\documentclass[amsmath,amssymb,eqsecnum,showpacs,nofootinbib,twocolumn]{revtex4-1}
\usepackage{chngcntr}

\usepackage{graphicx}
\usepackage{bm}
\usepackage{dcolumn}
\usepackage{color}

\newcommand{\rc}[1]{}

\newcommand{\rt}{r}
\newcommand{\tht}{\theta}
\newcommand{\tit}{t}

\newcommand{\dc}{k}

\newcommand{\Qg}{g}

\newcommand{\RSET}[1]{  \langle #1 \rangle}

\newcommand{\CrtSL}{\rt_{\rm{SL}}}

\newcommand{\op}{\Omega_H}

\newcommand{\be}{\begin{equation}}
\newcommand{\ee}{\end{equation}}
\newcommand{\bea}{\begin{eqnarray}}
\newcommand{\eea}{\end{eqnarray}}

\begin{document}

\counterwithout{equation}{section}

\author{Marc Casals}
\email{mcasals@cbpf.br, marc.casals@ucd.ie.}
\affiliation{Centro Brasileiro de Pesquisas F\'isicas (CBPF), Rio de Janeiro, CEP 22290-180, Brazil,}
\affiliation{School of Mathematical Sciences and Complex \& Adaptive Systems Laboratory, University College Dublin, Belfield, Dublin 4, Ireland.}

\author{Alessandro Fabbri}
\email{afabbri@ific.uv.es.}
\affiliation{Centro Studi e Ricerche E. Fermi, Piazza del Viminale 1, 00184 Roma, Italy,}
\affiliation{Dipartimento di Fisica dell'Universit\`a di Bologna and INFN sezione di Bologna, Via Irnerio 46, 40126 Bologna, Italy,}
\affiliation{Laboratoire de Physique Th\'eorique, CNRS UMR 8627, B\^at. 210, Univ. Paris-Sud, Universit\'e  Paris-Saclay, 91405 Orsay Cedex, France,}
\affiliation{Departamento de F\'isica Te\'orica and IFIC, Universidad de Valencia-CSIC, C. Dr. Moliner 50, 46100 Burjassot, Spain.}

\author{Cristi\'an Mart\'{\i}nez}
\email{martinez@cecs.cl.}
\author{ Jorge Zanelli}
\email{z@cecs.cl.}
\affiliation{Centro de Estudios Cient\'{\i}ficos (CECs), Av. Arturo Prat 514, Valdivia 5110466, Chile.}

\title{Quantum Backreaction on Three-Dimensional Black Holes and Naked Singularities}

\begin{abstract} 
We analytically investigate  backreaction by a quantum scalar field on two rotating  Ba\~nados-Teitelboim-Zanelli (BTZ) geometries:  that of a black hole and that of a naked singularity. In the former case, we explore the quantum effects on various regions of  relevance for a rotating black hole space-time. We find that the quantum effects lead to a growth of  both the event horizon and the radius of the ergosphere, and to a reduction of the angular velocity, compared to the unperturbed values. Furthermore, they give rise to the formation of a curvature singularity at the Cauchy horizon and show no  evidence of the appearance of a superradiant instability. In the case of a naked singularity, we find that quantum effects lead to the formation of a horizon that shields it, thus  supporting evidence for the r\^ole of quantum mechanics as a cosmic censor in nature.
\end{abstract}

\date{\today}
\maketitle


\textbf{Introduction.} It is expected that most black holes possess some rotation (e.g.,~\cite{gammie2004black,PhysRevD.67.064025}). The geometrical structure of rotating black hole space-times is a lot richer than that of nonrotating black hole space-times. For example, in Kerr space-time [i.e., a ($3+1$)-dimensional, rotating and asymptotically flat black hole] there exists a region -- the ergosphere -- ``near" the event horizon where observers cannot remain static: they must rotate in the same direction as the black hole. There also exists a region separated from the event horizon where an observer corotating with the  horizon must have a velocity greater than or equal to the speed of light; the boundary of such a region is called the speed-of-light surface. Inside a Kerr black hole, there is also the so-called inner horizon, which is a Cauchy horizon for data ``outside" the black hole. Beyond the inner horizon in the inward direction there exist closed null and timelike geodesics. None of these regions (ergosphere, speed-of-light surface or inner horizon) exist in the nonrotating limit of the Kerr geometry, a Schwarzschild black hole space-time (although an inner horizon does exist for a charged spherically-symmetric --Reissner-Nordstr\"om-- black hole).

The presence of the above regions has important consequences for the physics of black holes, notably for their stability properties. For example, the inner horizon of the Reissner-Nordstr\"om solution is classically unstable~\cite{poisson1989inner,PRD41Israel-Poisson} (phenomenon of ``mass inflation"). A similar feature occurs in the Kerr geometry, where  perturbations falling into the black hole are expected to produce a divergent curvature  at the inner horizon~\cite{brady1995nonlinear,ori1992structure,dotti2012unstable}. The presence of the ergosphere in Kerr, in its turn,  leads to the Penrose process~\cite{Penrose}, with its ``collisional" variant~\cite{PhysRevLett.103.111102}, and to the phenomenon of superradiance \cite{starobinskii1973amplification,zel1971generation}, whereby  matter --particles in the first case and boson field waves in the second-- falling into the black hole may be used in order to extract rotational energy from a  black hole.

The speed-of-light surface also plays an important r\^ole in the existence of superradiant modes. Superradiance in general is the cause behind various classical instabilities of black holes: under massive linear field perturbations \cite{zouros1979instabilities}; when the black hole is surrounded by a mirror \cite{Press:1972zz} (the so-called ``black hole bomb") which encloses any part of the region outside the speed-of-light surface~\cite{duffy2008renormalized}, and when a black hole lies in an anti de Sitter (AdS) universe (i.e., a universe with a negative cosmological constant) and is sufficiently small so that there exists a speed-of-light surface~\cite{hawking1999charged,PhysRevD.70.084011,winstanley2001classical}. Quantum-mechanically, the existence  of a  speed-of-light surface seems~\cite{kay1991theorems,Ottewill:2000qh} to be the reason why one cannot define a state describing a rotating black hole in thermal equilibrium with its own quantum boson field radiation~\cite{hartle1976path,frolov1989renormalized}. One may define such quantum state, however, if one excludes the region of the space-time beyond the   speed-of-light surface by placing a mirror~\cite{duffy2008renormalized} or, possibly, and more naturally, by placing the rotating black hole in an AdS universe for a sufficiently large cosmological constant~\cite{winstanley2001classical}. We finally note that the ergosphere can be present in a rotating space-time without an event horizon (such as that of a star), in which case it  leads to classical instabilities of the space-time~\cite{friedman1978ergosphere} and to quantum (Starobinski\u{\i}-Unruh) radiation~\cite{Unruh:1974bw}.

It is of great interest to understand the fate of the stability properties and rotating space-time regions in the presence of quantum corrections. One possibility is to study the backreaction effects from quantum matter on these geometrical regions of a rotating space-time. While such a  study would be technically very difficult in a Kerr black hole space-time (whether or not placed inside a mirror or a AdS universe), in this Letter we undertake that study  for a rotating black hole in $(2+1)$-dimensions, the so-called rotating BTZ (Ba\~nados-Teitelboim-Zanelli) black hole~\cite{banados1992black}. This black hole possesses inner (Cauchy) and outer (event) horizons and an ergosphere but no speed-of-light surface; its Cauchy horizon exhibits ``mass inflation"~\cite{chan1996interior,husain1994radiation};  its ergosphere leads to a Penrose-like process~\cite{cruz1994geodesic}. A major simplification in ($2+1$)-dimensions is the absence of propagating gravitational degrees of freedom, which eliminates the need for quantizing the gravitational field. Therefore, all quantum corrections come from the ``matter sector".

In this Letter, we analytically solve the semiclassical Einstein equations  sourced by a  conformally coupled and massless quantum scalar field on a rotating BTZ space-time. We  obtain the quantum-backreacted metric and investigate the quantum effects on the inner horizon,  outer horizon and   ergosphere, and investigate the possible creation  of  a speed-of-light surface. To the best of our knowledge, this is the first time that a quantum-backreacted metric has been obtained for a {\it rotating} black hole space-time\footnote{In the nonrotating case, the quantum-backreacted metric has been obtained for a background BTZ black hole in~\cite{martinez1997back} and for a background naked singularity space-time in~\cite{casals2016quantum}.}. Our results not only provide an insight into  backreaction effects that might take place for astrophysical  or particle-collider, rotating black holes,  but they can also be of interest for the AdS/Conformal Field Theory (CFT) conjectured correspondence (e.g.,~\cite{BSS}). For example, they can be used, following the analysis in~\cite{efk}, to test AdS/CFT in Randall Sundrum braneworlds \cite{rs2} by constructing a rotating black hole localised on an $\text{AdS}_3$ brane embedded in $\text{AdS}_4$~\cite{ehm}.

Most of the construction needed to analyze the backreaction of the black hole geometry produced by quantum fields can also be used to study the fate of a naked singularity. This could shed light on Penrose's Cosmic Censorship Hypothesis~\cite{Penrose}, which in its weaker version essentially states that, generically, no ``naked" (i.e., not covered by an event horizon) space-time singularities can form in Nature. In~\cite{casals2016quantum}  we  showed that quantum effects on a static naked singularity in $AdS_3$ lead to the formation of a horizon that covers it, thus enforcing Cosmic Censorship. In this Letter we confirm that quantum corrections continue to provide a mechanism for Cosmic Censorship in the spinning case as well. Therefore, this is not a peculiar feature of the static geometry, but a more generic phenomenon.


\textbf{Rotating BTZ geometry.}
The rotating BTZ geometry is obtained by identifying points in the universal covering of anti-de Sitter space-time (CAdS$_3$) by some spacelike Killing vector field corresponding to a  generator of certain global isometries of $\text{AdS}_3$.  Any open set of this geometry is, therefore, indistinguishable from a portion of 
$\text{CAdS}_3$.  The BTZ metric is given by \cite{banados1992black,banados1993geometry}
\begin{equation} \label{eq:RBTZ}
ds^2= \left(M-\frac{r^2}{\ell^2}\right)dt^2 -Jdt d\tht +\frac{dr^2}{\frac{r^2}{\ell^2}-M+\frac{J^2}{4\rt^2}}+r^2d\tht^2,
\end{equation}
where $t \in (-\infty,+\infty)$, $r \in (0, \infty)$ and $\tht \in [0,2\pi)$ and the  cosmological constant is given by $\Lambda = -\ell^{-2}$. The BTZ geometry corresponds to either a black hole or to a naked singularity possessing\footnote{We are choosing units such that the gravitational constant is $G=1/8$  and the three-dimensional Planck's length is $l_P=\hbar G$.} mass  $M$  and angular momentum $J$.
The metric (\ref{eq:RBTZ}) is stationary and axially symmetric, with corresponding Killing vectors $\xi\equiv \partial/\partial\tit$ and $\psi\equiv\partial/\partial\tht$, respectively.

In the case of the rotating BTZ black hole, $M\ell \geq |J|$ (the extremal case corresponding to the equality), the identification Killing field is a noncompact spacelike field -- see Eq.(\ref{Kv}) below. The resulting black hole space-time possesses an inner (Cauchy) horizon at $r=r_-= \ell |\alpha_{-}|/2$ and an outer (event) horizon at $r=r_+=\ell \alpha_{+}/2 $, where 
\begin{equation} \label{alfa+-}
\alpha_{\pm} \equiv \sqrt{M+\frac{J}{\ell}}\pm\sqrt{M-\frac{J}{\ell}} \, .
\end{equation}
The inner horizon is classically unstable~\cite{chan1996interior} in a similar manner to that of Kerr or Reissner-Nordstr\"om space-times~\cite{poisson1989inner,brady1995nonlinear,ori1992structure}. Unlike Kerr, the 2+1 black hole possesses no curvature singularities but it does possess a causal singularity at $\rt=0$: there are inextendible incomplete geodesics that hit $\rt=0$ \cite{banados1993geometry}.
The Killing vector $\xi$ is timelike for $r> \CrtSL \equiv\sqrt{M}\ell$, is null at $r=\CrtSL$ and is spacelike for $r \in(r_+,\CrtSL)$. This means that no static observers can exist for $r<\CrtSL$. The hypersurface $r=\CrtSL$ is hence called the static limit surface and the region $r \in (r_+,\CrtSL)$ is called the ergosphere. In its turn, the Killing vector $\chi\equiv \xi+\op \psi$, where $\op=J/(2r_+^2)$ is the angular velocity of the event horizon, is the  generator of the event horizon. The vector $\chi$ is null at the event horizon and, in the nonextremal case, is timelike everywhere outside. This means that, in the nonextremal case, observers that rigidly rotate at the angular velocity of the black hole can exist anywhere outside the event horizon, i.e., there is no speed-of-light surface. In the extremal case, on the other hand, the  Killing vector $\chi$ is null everywhere on and outside the event horizon.

For $M \ell \leq -|J|$ the metric (\ref{eq:RBTZ}) describes a conical singularity, also obtained by an identification in CAdS$_3$ by a spacelike Killing vector, which in this case is compact. Note that Eq.(\ref{alfa+-}) implies that, in this case, $\alpha_{\pm}$ are both purely imaginary: no horizon is present and the geometry is a true naked singularity. 
In this geometry, $\xi$ is always timelike and so there is no ergosphere.
The extremal case corresponds to maximal rotation,  $M\ell=-|J|$.

Finally, we note that in the non extremal cases, $|M|\ell > |J|$, the classical solutions can be obtained by boosting the corresponding static ($J=0$) black hole \cite{MTZ} or conical solution \cite{MZ}.


\textbf{Backreacted geometry.} The backreaction of quantum matter onto the geometry can be calculated via the \textit{semiclassical} Einstein equations:
\begin{equation} \label{semi}
G_{\mu \nu} - \ell^{-2} g_{\mu \nu}= \pi  \RSET{T_{\mu  \nu}}.
\end{equation}
Here, $G_{\mu \nu}$ is the Einstein tensor for the quantum-backreacted metric $g_{\mu \nu}$ and $\RSET{T_{\mu  \nu}}$ is the renormalized expectation value of the stress-energy tensor (RSET) of the matter field in some quantum state. The quantum state is determined by imposing boundary conditions for the field on the AdS boundary: the timelike hypersurface $r=\infty$. We note that the RSET is calculated on the classical space-time, rather than on the quantum-backreacted one (with metric $g_{\mu \nu}$).

We shall consider a conformally coupled and massless scalar field satisfying ``transparent" boundary conditions on the AdS boundary. Transparent boundary conditions correspond to decomposing the scalar field using modes which are smooth on the entire Einstein static universe \cite{avis1978quantum,Lifschytz:1993eb}. We calculate the vacuum expectation value for the RSET of the scalar field in a state corresponding to transparent boundary conditions in the following way. We consider the BTZ geometry as obtained from the appropriate identification of points in $\text{CAdS}_3$ under  an element of the Lorentz group. We then apply the method of images to find the two-point function of the field equation in the BTZ geometry from that in $\text{CAdS}_3$ with the appropriate identification. We then obtain the RSET~\cite{RotatingBTZLong} from the two-point function in the standard way.

We choose the following form for a general, stationary and axisymmetric metric
\begin{equation}
\label{eq:ansatz}
ds^2= (-e^{2a}b + r^2 k^2)dt^2+ 2r^2 k dt d\tht +\frac{dr^2}{b} + r^2d\tht^2\ ,
\end{equation}
for some functions $a(r)$, $b(r)$, and $k(r)$, which are given by their classical values plus corrections of order $O(l_P)$, denoted by $a_1$, $b_1$ and $k_1$ respectively. The (potential) horizons are determined by the zeros of $b(r)$ which, to order $l_P$, is  $b(\rt)= \left(\rt^2/\ell^2\right)-M+\left(J^2/(4\rt^2)\right)+l_P b_1(r)$. Next, we solve the semiclassical Einstein equations (\ref{semi}): in the left hand side, we insert the metric ansatz Eq.(\ref{eq:ansatz}) and expand to $O(l_P)$;  in the right hand side, we insert the RSET derived as indicated above. In order to integrate Einstein's equations, we fix the coordinate choice so that the values at infinity of the (rescaled) lapse and shift functions are equal to, respectively, 1 and 0, following the choice made in~\cite{banados1993geometry} for the classical unperturbed metric. The remaining two integration constants are the mass $M$ and angular momentum $J$, which, in order to make a significant comparison, we assume to have the same values as in the unperturbed solution.
 In this way, we find analytic expressions for $a_1$, $b_1$ and $k_1$, which we give elsewhere~\cite{RotatingBTZLong}.
In particular,
we find that, at large distances, the quantum corrections decay as: $a_1 = O(r^{-3}),\ b_1= O(r^{-1}),\ k_1= O(r^{-3})$. 
In the static limit ($J=0$, $\alpha_-=0, \alpha_+=2\sqrt{M}$), we recover the known results \cite{martinez1997back}:  $a=0$, $b_1=O(1/r)$, $ k=0$. 
Specifically,
we find the metric coefficient $b_1(r)$ to be of the form~\cite{RotatingBTZLong}:
\begin{equation}\label{b1}
b_1 = -\sum_{n=1}^N \frac{F_n(r)}{d_n(r)^{3/2}}\, ,
\end{equation} 
where $N=\infty$ in the black hole case and is finite in the naked singularity case.
Here, $F_n(r)$ is a function that for large $r$ grows as $r^2$ and $d_n(r)$ is the squared geodesic distance between a point and its $n$th image under the identification in CAdS$_3$; it can be written as  $d_n(r)=D_n r^2+E_n$ for some coefficients $D_n$ and $E_n$~\cite{RotatingBTZLong}.

\textbf{Horizons and other regions of the black hole geometry.}
We next investigate  various geometrical regions of interest of the backreacted rotating black hole metric.
For the black hole, the upper summation bound $N=\infty$ and for fixed $r>r_-$, where $d_n(r)>0$, this is a converging geometric series~\cite{steif1994quantum}. The metric perturbations diverge for $d_n(r_n)=0$, which occurs for certain discrete radii satisfying $0<r_n<r_-$. As $n\to +\infty$, $r_n\to r_-$, and therefore the inner horizon becomes a surface with an accumulation of points where $d_n=0$. This is a direct consequence of the identification that produces the spinning black hole. The Killing vector that is employed in this identification is\footnote{The $SO(2,2)$ generators are $J_{ab}=x_a\partial_b-x_b\partial_a$, see e.g., \cite{banados1993geometry}.}
\begin{equation} \label{Kv}
\zeta(r_+,r_-)=r_+ J_{12} - r_- J_{03},
\end{equation}
whose norm, $\zeta\cdot \zeta = r_+^2 - r_-^2$, is positive for a nonextremal black hole. The spacelike vector $\zeta(r_+,r_-)$, however, can identify two distinct points connected by a null geodesic in the CAdS$_3$, turning this curve into a closed null solution of  the geodesic equation in the black hole geometry.  (We note that this curve is not everywhere future directed, as opposed to the closed timelike curves that would be produced if the identification in CAdS$_3$ was made with a timelike Killing vector: those curves would be everywhere future directed or everywhere past directed). The resulting null closed curve extends from infinity to some radius $r_{\text{min}}$ inside the inner horizon and back to infinity. This means that this geodesic is not a serious issue in classical physics because no real massless particle can follow this trajectory crossing both horizons twice~\cite{cruz1994geodesic,ACZ}. Virtual quantum mechanical particles, however, do not respect causality  and we find that this gives a divergent contribution to the RSET coming from a pole in the propagator at a series of circles approaching $r_-$ from the inside. This accumulation of poles produces an essential singularity at $r_-$. We 
find~\cite{RotatingBTZLong} that the Kretschmann invariant picks up a divergent contribution proportional to the square of  the RSET. Therefore, the geometry indeed develops a curvature singularity at $r_-$ and, consequently, the semiclassical approximation can only be trusted for $r>r_-$.
Although both the quantum backreaction found here and the classical mass inflation found in~\cite{chan1996interior,husain1994radiation} yield a diverging (local) stress-energy at $r_-$, this singularity is not of the same nature in the two cases (in mass inflation, it is due to infinitely-blueshifted perturbations generated in the external region); plus, here the Kretschmann scalar diverges whereas for mass inflation in BTZ it does not.


The backreacted radius of the outer horizon is given by the largest positive root of $b(r)=0$. Working at $O(l_P)$, the corrected event horizon radius (in the nonextremal case) is of the form $r^{(q)}_+= r_+ \left(1+l_P x_+\right)$, where 
\begin{equation}\label{eq:x+}
x_+\equiv -\frac{2b_1(r_+)}{\alpha_+^2 - \alpha_-^2},
\end{equation}
and $b_1(r_+)$ is negative. Therefore, the event horizon grows, $r_+^{(q)}>r_+$. We note that the expression for $r^{(q)}_+$ via Eq.(\ref{eq:x+}) is only valid for $l_P\ll (r_+-r_-)$. In the opposite regime, $0<(r_+-r_-)\ll l_P$, the correction to the horizon radius has an expression different from Eq.(\ref{eq:x+}) \cite{RotatingBTZLong}. 
In the extremal case, $r_+=r_-$, and for $r_+^2\gg \ell^2 l_P b_1(r_+)$, this expression takes the form $r_+^{(q)}= r_+(1+ \sqrt{l_P} y_+)$, where
\begin{equation}
y_+\equiv  \sqrt{\frac{-b_1(r_+)}{2 M}}
\end{equation}
and
\begin{equation}
b_1(r_+)=-\frac{1}{\ell \pi^2}\sum_{n=1}^{\infty}\frac{1}{n^2 \sinh\left( \frac{n \pi \alpha_+}{2} \right)} \ .
\end{equation}
This limit coincides with the corrected $r_+^{(q)}$ for the extremal solution in the semiclassical approximation.

To find the boundary of the quantum-corrected ergosphere we need to solve $ \Qg_{tt}=-e^{2a(\rt)}b(\rt)+\rt^2\dc^2(\rt)=0$, which we solve to $O(l_P)$. We analytically find that the sign of the quantum correction to the radius of the  static limit surface is always positive.

We can also compute the quantum-corrected angular velocity of the black hole: $\Omega_H^{(q)}= \left. -\frac{\Qg_{\tit\tht}}{\Qg_{\tht\tht}}\right|_{\rt_+} = -\dc(\rt_+)$. We find numerical evidence that $\Omega_H-\Omega_H^{(q)}$ is always positive. We now turn to investigating the speed-of-light surface. The Killing vector $\chi^{(q)\mu}=\xi^\mu + \Omega_H^{(q)}\psi^\mu=(1,0,\Omega_H^{(q)})$ has norm $\Qg_{\mu\nu}\chi^{(q)\mu}\chi^{(q)\nu} = -e^{2a}b+\rt^2\dc^2+2\rt^2\dc\Omega_H^{(q)}+\rt^2\Omega_H^{(q)^2}$. The Killing vector $\chi^{(q)\mu}$, in the nonextremal case, is timelike in the near-horizon region and becomes null on the horizon. Near infinity we find that $\chi^{(q)2}\sim -\frac{\rt^2}{\ell^2}\left(1 -\ell^2\Omega_H^{(q)2}\right )$. The condition for $\chi^{(q)}$ to be spacelike, and (likely) for the space-time to develop a superradiant instability is $\ell\Omega_H^{(q)}=\ell\left(\frac{J}{2\rt_{+}^{(q)^2}}-l_P\dc_1(\rt_+)\right)>1$. We find that $\ell \Omega_H^{(q)}<\ell \Omega_H\leq 1$ (the equality being realized in the extremal case). We conclude that the quantum effects do not appear to change the superradiant-stability property of the classical solutions.

Another important (and delicate) case is the extremal limit $\alpha_+\to\alpha_-$.  In this case, the identification that yields the extremal black hole, $\zeta_{ext}=r_+(J_{01}-J_{23})+ (J_{12}+J_{03}+J_{02}-J_{13})/2$, is not obtained as the limit $r_+ \to r_-$ of (\ref{Kv}) and therefore it is not immediately obvious what happens in this case. However, we obtain that the extremal limit of the RSET in the nonextremal black hole is equal to the RSET in the extremal black hole; the backreacted metrics share the same feature. Therefore, our results are physically meaningful for nonextremal black holes all the way down to the extremal limit.

\textbf{Naked singularity and Cosmic Censorship.} 
For the nonextremal conical singularity ($M\ell<-|J| \leq 0$), the upper summation bound $N$ is finite and, therefore, convergence is not an issue. In this case, the quantum correction $b_1$ possesses at least one pole at a finite radius where $b_1 \to -\infty$. This implies that the quantum corrections always generate an event horizon that covers the conical singularity at $r=0$.
We note, however, that for finite values of $(M, J)$, the formed horizon has size $O(l_p)$ and our results are at most indicative (higher-order quantum corrections are equally important for establishing its presence). Instead, for masses just below $M=0$, and as in the static case, $r_+=O(l_p^{1/3})\gg O(l_p)$ appears to be physically meaningful~\cite{casals2016quantum}.
An alternative way of seeing this is by noting that the metric components as well as the corrections for $a$, $b$ and $k$ are  continuous and analytic in the $M$-$J$ plane for $|J|<|M|\ell$. Since in the static case the  quantum corrections give rise to a horizon at finite radius~\cite{casals2016quantum}, the addition of angular momentum produces a continuous change in this radius, and therefore, Cosmic Censorship continues to be upheld when angular momentum is switched on \cite{RotatingBTZLong}.


\textbf{Discussion.} We have established that the presence of a conformally coupled quantum scalar field on a rotating BTZ black hole leads to: (1) the event horizon growing ($r_+^{(q)} > r_+$), (2) the radius of the static limit surface growing ($\CrtSL^{(q)}>\CrtSL$),  (3) the angular velocity diminishing ($\Omega_H^{(q)}<\Omega_H$), and (4) no evidence that a speed-of-light surface forms.
In particular, in the extremal case, 
the generator of the  horizon goes from being null to timelike
everywhere outside the horizon, and so, in a sense, ``the quantum  corrections take the solutions away from extremality".

The perturbative correction shows the formation of a singularity at the inner horizon, which can be interpreted as an instability due to the existence of a curvature singularity there. In the extreme case, the event horizon also grows and the curvature singularity still forms inside, so that the black hole can no longer be called ``extremal".

Strictly speaking, however, the instability at $r_-$ signals a breakdown of the linear approximation itself, and therefore, any statement about the fate of the geometry there can be viewed, at most, as an indicative suggestion. 

Nevertheless, it can also be argued that the singularity of the RSET is not a perturbative  approximation but an exact result due to the existence of closed null curves (which are not everywhere future directed or past directed) in the background geometry. Therefore, the formation of a barrier of infinite energy is a real issue that cannot be dismissed on the grounds that the right hand side of (\ref{semi}) blows up, even if this equation could not provide an expression for the metric in the neighborhood of $r_-$. As a parallel, we note that, in the case of Kerr, Ref.~\cite{dotti2012unstable} directly links a classical instability of the region $r<r_-$ under linear field perturbations to the existence of
 closed-timelike curves in that region.

The only sure way of learning about the space-time geometry near the inner horizon would be to solve the coupled system (\ref{semi}) exactly, in which the two-point function and the RSET are computed in the corrected geometry. In the absence of such a scheme, the best one could achieve is a perturbative procedure where the corrected $\RSET{T^{(1)}_{\mu  \nu}}$ is the input to obtain a first corrected metric, $g^{(1)}_{\mu \nu}$, using (\ref{semi}). Next, this metric could be used to compute a new corrected stress-energy tensor, $\RSET{T^{(2)}_{\mu  \nu}}$, etc.

In the iterative procedure outlined above, there is no need to worry about quantum gravity effects for there are no gravitons in 2+1 dimensions, and therefore, there are no gravitational loop corrections. This is a significant difference with respect to the 3+1 case, where quantum gravity corrections cannot be consistently ignored. 

For $\ell M<-|J|$ (and possibly for $\ell M=-|J|$ as well), the classical naked singularity dresses up with a  horizon produced by quantum effects, as in the static case. We conclude that quantum mechanics provides a mechanism for Cosmic Censorship for spinning as well as for static conical singularities. 

\textbf{Acknowledgments.} M.C. acknowledges partial financial support by CNPq (Brazil), process number 308556/2014-3. A.F. acknowledges partial financial support from the Spanish MINECO through the grant FIS2014-57387-C3-1-P and the Severo Ochoa Excellence Center Project SEV-2014-0398. This work has been partially funded through Grants No. 1130658, No. 1140155, and No. 1161311 from FONDECYT. The Centro de Estudios Cient\'{\i}ficos (CECs) is funded by the Chilean Government through the Centers of Excellence Base Financing Program of CONICYT.

\begin{thebibliography}{38}
\expandafter\ifx\csname natexlab\endcsname\relax\def\natexlab#1{#1}\fi
\expandafter\ifx\csname bibnamefont\endcsname\relax
  \def\bibnamefont#1{#1}\fi
\expandafter\ifx\csname bibfnamefont\endcsname\relax
  \def\bibfnamefont#1{#1}\fi
\expandafter\ifx\csname citenamefont\endcsname\relax
  \def\citenamefont#1{#1}\fi
\expandafter\ifx\csname url\endcsname\relax
  \def\url#1{\texttt{#1}}\fi
\expandafter\ifx\csname urlprefix\endcsname\relax\def\urlprefix{URL }\fi
\providecommand{\bibinfo}[2]{#2}
\providecommand{\eprint}[2][]{\url{#2}}

\bibitem{gammie2004black}  C.~F.~Gammie, S.~L.~Shapiro and J.~C.~McKinney, Astrophys.\ J.\  {\bf 602}, 312 (2004) 
[astro-ph/0310886].


\bibitem{PhysRevD.67.064025} D.~Ida, K.~Y.~Oda and S.~C.~Park, Phys.\ Rev.\ D {\bf 67}, 064025 (2003); Erratum: [Phys.\ Rev.\ D {\bf 69}, 049901 (2004)] 
[hep-th/0212108].

\bibitem{poisson1989inner} E.~Poisson and W.~Israel, Phys.\ Rev.\ Lett.\  {\bf 63}, 1663 (1989). 

\bibitem{PRD41Israel-Poisson}E.~Poisson and W.~Israel, Phys.\ Rev.\ D {\bf 41}, 1796 (1990).

\bibitem{brady1995nonlinear} P.~R.~Brady and C.~M.~Chambers, Phys.\ Rev.\ D {\bf 51}, 4177 (1995) [gr-qc/9501025].

\bibitem{ori1992structure}A.~Ori, Phys.\ Rev.\ Lett.\ {\bf 68}, 2117 (1992). 

  
  

\bibitem{dotti2012unstable} G.~Dotti, R.~J.~Gleiser and I.~F.~Ranea-Sandoval, Class.\ Quant.\ Grav.\  {\bf 29}, 095017 (2012)
[arXiv:1111.6854 [gr-qc]].
  
\bibitem{Penrose} R. Penrose, in Black Holes and Relativistic Stars, edited by Robert Wald (University of Chicago Press, Chicago 1998), Chap 5;
J. Astrophys. Astr. \textbf{20}, 233 (1999);
Riv. Nuovo Cimento \textbf{1}, 252 (1969).
  

\bibitem{PhysRevLett.103.111102} M.~Ba\~nados, J.~Silk and S.~M.~West, Phys.\ Rev.\ Lett.\  {\bf 103}, 111102 (2009)
[arXiv:0909.0169 [hep-ph]].

\bibitem{starobinskii1973amplification}A.~Starobinskii, Zh. Eksp. Teor. Fiz, \textbf{64}, 48 (1973) [Sov. Phys. JETP {\bf 37}, 28 (1973)].

\bibitem{zel1971generation} Y.~B.~Zel'Dovich, ZhETF Pisma Redaktsiiu \textbf{14}, 270 (1971) [JETP Lett. \textbf{14}, 270 (1971)].

  

\bibitem{zouros1979instabilities} T.~J.~M.~Zouros and D.~M.~Eardley, Ann. Phys. (N.Y.)\  {\bf 118}, 139 (1979).

\bibitem{Press:1972zz}W.~H.~Press and S.~A.~Teukolsky, Nature (London) \textbf{238}, 211 (1972).

\bibitem{duffy2008renormalized} G.~Duffy and A.~C.~Ottewill, Phys.\ Rev.\ D {\bf 77}, 024007 (2008)
[gr-qc/0507116].

\bibitem{hawking1999charged}S.~W.~Hawking and H.~S.~Reall, Phys.\ Rev.\ D {\bf 61}, 024014 (1999)
[hep-th/9908109].

\bibitem{winstanley2001classical}E.~Winstanley, Phys.\ Rev.\ D {\bf 64}, 104010 (2001)
[hep-th/9908109].

\bibitem{PhysRevD.70.084011} V.~Cardoso and O.~J.~C.~Dias, Phys.\ Rev.\ D {\bf 70}, 084011 (2004) 
[hep-th/0405006].

\bibitem{kay1991theorems}B.~S.~Kay and R.~M.~Wald, Phys.\ Rept.\  {\bf 207}, 49 (1991).

\bibitem{Ottewill:2000qh}A.~C.~Ottewill and E.~Winstanley, Phys.\ Rev.\ D {\bf 62}, 084018 (2000)
[gr-qc/0004022].
  
\bibitem{hartle1976path}J.~B.~Hartle and S.~W.~Hawking, Phys.\ Rev.\ D {\bf 13}, 2188 (1976).

\bibitem{frolov1989renormalized}V.~P.~Frolov and K.~S.~Thorne, Phys.\ Rev.\ D {\bf 39}, 2125 (1989).

\bibitem{friedman1978ergosphere} J.~L.~Friedman, Commun.\ Math.\ Phys.\  {\bf 63}, No. 3, 243 (1978). 

\bibitem{Unruh:1974bw}W.~G.~Unruh, Phys.\ Rev.\ D {\bf 10}, 3194 (1974).

\bibitem{banados1992black}M.~Ba\~nados, C.~Teitelboim and J.~Zanelli, Phys.\ Rev.\ Lett.\  {\bf 69}, 1849 (1992) [hep-th/9204099].

\bibitem{chan1996interior} J.~S.~F.~Chan, K.~C.~K.~Chan and R.~B.~Mann, Phys.\ Rev.\ D {\bf 54}, 1535 (1996) [gr-qc/9406049].

\bibitem{husain1994radiation} V.~Husain, Phys.\ Rev.\ D {\bf 50}, R2361 (1994) [gr-qc/9404047].

\bibitem{cruz1994geodesic} N.~Cruz, C.~Mart\'{\i}nez and L.~Pe\~na, Class.\ Quant.\ Grav.\  {\bf 11}, 2731 (1994) [gr-qc/9401025].
  
\bibitem{martinez1997back}C.~Mart\'{\i}nez and J.~Zanelli, Phys.\ Rev.\ D {\bf 55}, 3642 (1997) [gr-qc/9610050].

\bibitem{casals2016quantum} M.~Casals, A.~Fabbri, C.~Mart\'{\i}nez and J.~Zanelli, Phys.\ Lett.\ B {\bf 760}, 244 (2016) [arXiv:1605.06078 [hep-th]]. 

%
\bibitem{BSS} D.~Birmingham, I.~Sachs and S.~N.~Solodukhin,  Phys.\ Rev.\ Lett. {\bf 88}, 151301 (2002).

\bibitem{efk} R.~Emparan, A.~Fabbri and N.~Kaloper, J. High Energy Phys.  {\bf 08} (2002) 043.

\bibitem{rs2} L.~Randall and  R.~Sundrum, Phys. Rev. Lett. {\bf 83}, 4690 (1999).   

\bibitem{ehm} R.~Emparan, G.T.~Horowitz and R.C.~Myers, JHEP {\bf 0001}, 007 (2000); JHEP {\bf 0001}, 021 (2000).



\bibitem{banados1993geometry} M.~Ba\~nados, M.~Henneaux, C.~Teitelboim and J.~Zanelli, Phys.\ Rev.\ D {\bf 48}, 1506 (1993); Erratum: [Phys.\ Rev.\ D {\bf 88}, 069902 (2013)] [gr-qc/9302012].
 
\bibitem{MTZ} C.~Mart\'{\i}nez, C.~Teitelboim and J.~Zanelli, Phys.\ Rev.\ D {\bf 61}, 104013 (2000) [hep-th/9912259].
 
\bibitem{MZ} O.~Mi\v{s}kovi\'c and J.~Zanelli, Phys.\ Rev.\ D {\bf 79}, 105011 (2009) [arXiv:0904.0475 [hep-th]].
 
\bibitem{avis1978quantum} S.~J.~Avis, C.~J.~Isham and D.~Storey, Phys.\ Rev.\ D {\bf 18}, 3565 (1978). 
  
\bibitem{Lifschytz:1993eb} G.~Lifschytz and M.~Ortiz, Phys.\ Rev.\ D {\bf 49}, 1929 (1994) [gr-qc/9310008].

\bibitem[{\citenamefont{Rotating BTZ Long}(2016)\citenamefont{Casals, Fabbri,
  Mart{\'\i}nez, and Zanelli}}]{RotatingBTZLong}
\bibinfo{author}{\bibfnamefont{M.}~\bibnamefont{Casals}},
  \bibinfo{author}{\bibfnamefont{A.}~\bibnamefont{Fabbri}},
  \bibinfo{author}{\bibfnamefont{C.}~\bibnamefont{Mart{\'\i}nez}},
  \bibnamefont{and} \bibinfo{author}{\bibfnamefont{J.}~\bibnamefont{Zanelli}},
 \bibinfo{note}{in preparation}   (\bibinfo{year}{2017}).
 
\bibitem{steif1994quantum} A.~R.~Steif, Phys.\ Rev.\ D {\bf 49}, R585 (1994) [gr-qc/9308032].

\bibitem{ACZ} C.~T.~Asplund, N.~Callebaut, C.~Zukowski,
J. High Energy Phys. 09 (2016) 154
 [arXiv:1604.02687 [hep-th]].



\end{thebibliography}

\end{document}